\documentclass[floatfix,aps,prl,showpacs,amsmath,nofootinbib,preprintnumbers,preprint]{revtex4}
\usepackage{graphicx}
\usepackage{epstopdf}
\usepackage{amsmath}
\usepackage{amsfonts}
\usepackage{amssymb}
\usepackage{color}
\usepackage{mathrsfs}
\usepackage{multirow}
\usepackage{bm}

\def\({\left(}
\def\){\right)}
\def\[{\left[}
\def\]{\right]}

\def\e{\begin{equation}}
\def\q{\end{equation}}
\def\m{\begin{eqnarray}}
\def\n{\end{eqnarray}}

\begin{document}

\title{No evidence for the blue-tilted power spectrum of relic gravitational waves}

\author{Qing-Guo Huang}\email{huangqg@itp.ac.cn}
\author{Sai Wang}\email{wangsai@itp.ac.cn}
\affiliation{State Key Laboratory of Theoretical Physics, Institute of Theoretical Physics, Chinese Academy of Science, Beijing 100190, China}

\date{\today}

\begin{abstract}

In this paper, we constrain the tilt of the power spectrum of relic gravitational waves by combining the data from BICEP2/Keck array and Planck (BKP) and the Laser Interferometer Gravitational-Waves Observatory (LIGO). From the data of BKP B-modes, the constraint on the tensor tilt is $n_t=0.66^{+1.83}_{-1.44}$ at the $68\%$ confidence level. By further adding the LIGO upper limit on the energy density of gravitational waves, the constraint becomes $n_t=-0.76^{+1.37}_{-0.52}$ at the $68\%$ confidence level. We conclude that there is no evidence for a blue-tilted power spectrum of relic gravitational waves and either sign of the index of tensor power spectrum is compatible with the data.

\end{abstract}


\maketitle

%%%%%%%%%%%%%%%%%%%%%%%%%%%%%%%%%%%%%%%%
%%%%%%%%%%%%%%%%%%%%%%%%%%%%%%%%%%%%%%%

The primordial gravitational waves can be generated during inflation \cite{Starobinsky:1979,Starobinsky:1980te,Guth:1980zm,Linde:1981mu,Albrecht:1982wi}. The simplest inflation models show the power spectrum of primordial gravitational waves to be adiabatic, Gaussian and nearly scale-invariant. The primordial gravitational waves contribute to both the total intensity and polarizations for the anisotropy of the cosmic microwave background (CMB) \cite{Grishchuk:1974ny,Starobinsky:1979ty,Rubakov:1982,Crittenden:1993ni,Kamionkowski:1996zd,Kamionkowski:1996ks,Hu:1997mn}. Especially, the effects of primordial gravitational waves on the B-mode polarization might be detectable at the range $\ell\lesssim150$. Recently, Background Imaging of Cosmic Extragalactic Polarization (BICEP2) \cite{Ade:2014xna} reported an excess of B-mode power over the base lensed-$\Lambda$CDM expectation at the range $30<\ell<150$ with a significance of more than $5\sigma$. However, there were debates on whether the B-mode signals are resulted from the primordial gravitational waves or from the interstellar dust polarization \cite{Mortonson:2014bja,Flauger:2014qra,Colley:2014nna,Cheng:2014pxa}. Actually, the power of dust polarization has the same magnitude as the BICEP2 B-mode signals \cite{Adam:2014oea}. Most recently, a joint analysis of the data from BICEP2/Keck array and Planck (BKP) \cite{Ade:2015tva} yielded an upper limit on the tensor-to-scalar ratio, i.e., $r_{0.05}<0.12$ at the $95\%$ confidence level (C.L.) which is compatible with the Planck temperature-only limit. The significance is too low to claim a detection of primordial gravitational waves.

The tilt of power spectrum of relic gravitational waves is one other significant quantity. It describes the shape of tensor power spectrum. In general, the tensor tilt $n_t$ is defined by
\begin{equation}
\label{nt}
P_t(k)=A_t \left(\frac{k}{k_p}\right)^{n_t}\ ,
\end{equation}
where $P_t(k)$ denotes the amplitude of the primordial tensor power spectrum at scale $k$, and $k_p=0.01\rm{Mpc}^{-1}$ is the pivot scale. In the canonical single-field slow-roll inflation models, the tensor tilt is determined by a consistency relation between $r$ and $n_t$, namely, $r=-8n_t$ \cite{Liddle:1992wi}. Because of the upper limit $r_{0.05}<0.12$, the spectrum of relic gravitational waves is nearly scale-invariant. In the inflation model, the tensor tilt is generally predicted as $n_t=-2\epsilon$ \cite{Liddle:1992wi,Garriga:1999vw}. The inflation requires $\ddot{a}/a=H^2(1-\epsilon)$ where $\epsilon=\dot{H}/H^2$, and thus $-2<n_t<0$. However, there are also certain alternatives of the inflation model, which predict different tensor tilt. For example, the ekpyrotic model \cite{Khoury:2001wf} predicts a blue-tilted tensor power spectrum, i.e. $n_t=2$. Here we take the tensor tilt $n_t$ as a fully free parameter in our analysis.

Besides constraining the tensor-to-scalar ratio, the current data can also constrain the tilt of the primordial tensor power spectrum. In this paper, we shall make a joint analysis of the CMB B-mode polarization and Laser Interferometer Gravitational wave Observatory (LIGO) \cite{Aasi:2014zwg} data to constrain the tensor tilt $n_t$. The CMB B-mode polarization data coming from a joint analysis of the data BICEP2/Keck array and Planck \cite{Ade:2015tva} can be used to  constrain the tilt of the power spectrum of relic gravitational waves at large angular scales. In this paper, we are just interested in the BB bandpowers $1-5$ which are taken between BICEP2/Keck and the 217 and 353 GHz bands of Planck. %Actually, the full 9 bandpowers give a quite similar constaint.
As a complement, the data of LIGO place an upper limit on the energy density of relic gravitational waves at a specific frequency band around $100\rm{Hz}$.

The data of LIGO refers to the upper limit on the energy density of relic gravitational waves at the frequency band around $100\rm{Hz}$. The upper limit is given by $\Omega_{GW}<5.6\times10^{-6}$ at the $95\%$ confidence \cite{Aasi:2014zwg}. In general, the energy density of relic gravitational waves at the wavenumber $k$ is determined by \cite{Turnersk1993,Zhao:2013bba}
\begin{equation}
\label{energy density of relic GW}
\Omega_{GW}(k)=\frac{P_t(k)}{12H_0^2}\left(\dot{\mathcal{T}}(\eta_0,k)\right)^2\ ,
\end{equation}
where $\eta_0$ denotes the conformal time today, $H_0$ is the Hubble parameter today, $\mathcal{T}(\eta,k)$ is the transfer function of tensor perturbations, and the overdot denotes a cosmic time derivative $d/dt$. The wavenumber $k$ of relic gravitational waves is related to the frequency $f$ by $f=k/2\pi$. The tensor transfer function $\mathcal{T}(\eta,k)$ describes the evolution of relic gravitational waves in the universe. It has an analytical approximation, namely, \cite{Turnersk1993,Kuroyanagisk2010,Zhao:2013bba,Watanabesk2006}
\begin{equation}
\label{tensor transfer function}
\dot{\mathcal{T}}(\eta_0,k)=\frac{-3j_2(k\eta_0)\Omega_m}{\eta_0}\sqrt{1+1.36\left(\frac{k}{k_{eq}}\right)+2.50\left(\frac{k}{k_{eq}}\right)^2}\ ,
\end{equation}
where $\eta_0=1.41\times10^4\rm{Mpc}$ is the conformal time today, and $k_{eq}=0.073\Omega_m h^2\rm{Mpc}^{-1}$ denotes the wavenumber relating to the Hubble horizon at the time of matter-radiation equality. Here $\Omega_m$ and $h$ denote the matter density parameter and the reduced Hubble parameter today, respectively. Since LIGO is sensitive to the relic gravitational waves of the wavenumber $k\gg k_{eq}$, the energy density of relic gravitational waves today is given by \cite{Zhao:2013bba}
\begin{equation}
\label{RGW energy density}
\Omega_{GW}(k)\simeq\frac{15}{16}\frac{\Omega_m^2A_s r}{H_0^2\eta_0^4 k_{eq}^2}\left(\frac{k}{k_p}\right)^{n_t}\ .
\end{equation}
The data of LIGO can be a complement to the CMB data, since both observations refer to two very different cosmological scales. Thus, one can make certain multi-wavelength constraints on the inflationary physics \cite{Meerburg:2015zua}.

We add a prior into the public Markov Chain Monte Carlo sampler (CosmoMC) \cite{Lewis:2002ah} to account for the LIGO upper limit on the energy density of relic gravitational waves. We consider two combined datasets: one is the BKP B-mode data; the other one refers to a combination of the BKP data and the LIGO upper limit (BKP+LIGO). We just consider the BKP B-mode data with the bandpowers $1-5$ ($20<\ell<200$), which are insensitive to the lensing effects. Actually, the extra four bandpowers would lead little changes on our final results. The cosmological model considered here is the base lensed-$\Lambda$CDM model$+$tensor cosmology. The parameters of the base lensed-$\Lambda$CDM model are fixed as the ``BKP cosmology parameters'' in the CosmoMC, namely,  $(\Omega_bh^2,\Omega_ch^2,100\theta_{MC},\tau,\ln(10^{10}A_s),n_s)=(0.0220323,0.1203761,1.0411,0.0924518,3.16,0.9619123)$. Thus, the parameter space just includes the tensor-to-scalar ratio ($r$), the tensor tilt ($n_t$).

\begin{figure}
\includegraphics[width=12 cm]{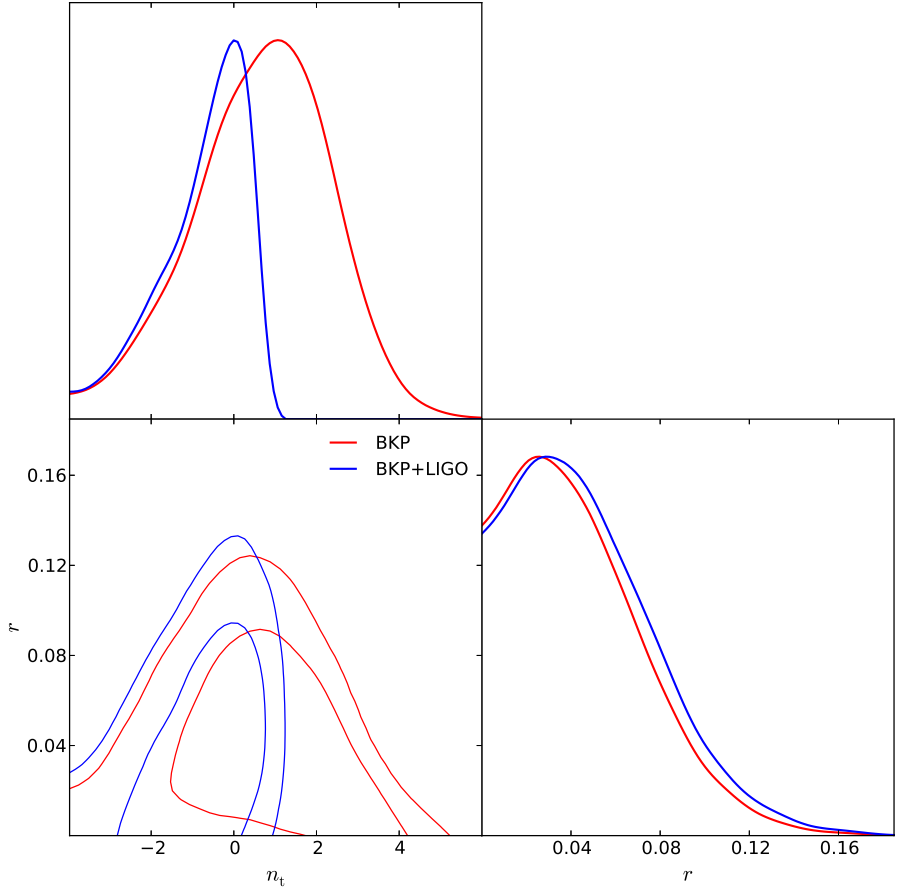}
\caption{The marginalized contour plot and likelihood distributions of the parameters $r$ and $n_t$ from the BKP only (red) and BKP+LIGO (blue) datasets. }
\label{figure:BKPLIGO}
\end{figure}

Our results are summarized in Table~\ref{table:fitting}.
\begin{table*}[!hts]
\footnotesize
\centering
\renewcommand{\arraystretch}{1.5}
\begin{tabular}{c|c|c|c|c}
\hline \hline
\multirow{2}{2cm}{parameter}& \multicolumn{2}{c|} {BKP}&\multicolumn{2}{c}{BKP+LIGO} \\
\cline{2-5}
 &$68\%$~C.L. &$95\%$~C.L. &$68\%$~C.L. &$95\%$~C.L.\\
\hline
$r$ &$<0.055$&$<0.099$ &$<0.059$ &$<0.106$  \\
$n_t$ &$0.66^{+1.83}_{-1.44}$ &$0.66^{+2.92}_{-3.42}$ &$-0.76^{+1.37}_{-0.52}$ &$-0.76^{+1.63}_{-2.21}$ \\
\hline
\end{tabular}
\caption{The $68\%$ and $95\%$ limits for the parameters $r$ and $n_t$ from the BKP only and BKP+LIGO datasets.}
\label{table:fitting}
\end{table*}
The marginalized contour plot and the likelihood distributions of the parameters $r$ and $n_t$ are illustrated in Fig.~\ref{figure:BKPLIGO}.
If the BKP dataset is considered, the constraints on the tensor-to-scalar ratio $r$ and the tensor spectral index $n_t$ are $r<0.099$ ($95\%$~C.L.) and $n_t=0.66^{+1.83}_{-1.44}$ ($68\%$~C.L.), respectively. If the BKP+LIGO dataset is considered, the constraints become $r<0.106$ ($95\%$~C.L.) and $n_t=-0.76^{+1.37}_{-0.52}$ ($68\%$~C.L.), respectively. Our results indicate that a scale-invariant power spectrum of relic gravitational waves is compatible with the data.
Furthermore, the ekpyrotic model in which $n_t=2$ is ruled out at much more than $3\sigma$ level once the LIGO dataset is taken into account.
The canonical single-field slow-roll inflation is within $1\sigma$. Note that the errorbar of $n_t$ is still too large to test the inflationary consistency relation.

In this paper, we make a constraint on the tilt of power spectrum of relic gravitational waves by combining the data of BKP and LIGO. The bounds on the tensor tilt are given by $n_t=0.66^{+1.83}_{-1.44}$ at $68\%$~C.L. for the BKP dataset, and $n_t=-0.76^{+1.37}_{-0.52}$ at $68\%$~C.L. for the BKP+LIGO dataset. We find that there is no evidence for the blue-tilted tensor power spectrum, and either sign of the tilt of tensor power spectrum is compatible with the data.
The LIGO upper limit on the energy density of relic gravitational waves gives a stringent upper bound on the tensor tilt. However, the lower bound on $n_t$ is still quite loose. Actually the temperature auto-correlations are sensitive to the negative value of the tilt of tensor power spectrum \cite{Cheng:2012je,Cheng:2014ota}. Thus, the constraint on the negative part of the tensor tilt will be significantly improved once the the CMB TT spectrum is taken into account. Even though we obtain certain constraints on the tensor tilt just from CMB B-mode and LIGO datasets, it is worthwhile to further combining other datasets, such as the Planck TT spectrum, to make a more stringent constraint on the tilt of tensor power spectrum in the near future.

\vspace{0.5cm}

\noindent {\bf Acknowledgments. }We acknowledge the use of ITP and Lenovo Shenteng 7000 supercomputer in the Supercomputing Center of CAS
for providing computing resources. S.W. thanks for useful discussion with Xin Li. This work is supported by the project of Knowledge Innovation Program of Chinese Academy of Science and grants from NSFC (grant NO. 11322545 and 11335012).

%%%%%%%%%%%%%%%%%%%%%%%%%%%%%%%%%%%%%%%%
%%%%%%%%%%%%%%%%%%%%%%%%%%%%%%%%%%%%%%%%

%%%%%%%%%%%%%%%%%%%%%%%%%%%%%%%%%%%%%%%%
%%%%%%%%%%%%%%%%%%%%%%%%%%%%%%%%%%%%%%%%

%%%%%%%%%%%%%%%%%%%%%%%%%%%%%%%%%%%%%%%%
%%%%%%%%%%%%%%%%%%%%%%%%%%%%%%%%%%%%%%%%
\end{document}